# DISTRIBUTED LANCE-WILLIAM CLUSTERING ALGORITHM

Yarmish, Gavriel, Brooklyn College, City University of New York
Listowsky, Philip, Kingsborough College, City University of New York
Dexter, Simon, City University of New York, Brooklyn College

Abstract: One important tool is the optimal clustering of data into useful categories. Dividing similar objects into a smaller number of clusters is of importance in many applications. These include search engines, monitoring of academic performance, biology and wireless networks. We first discuss a number of clustering methods. We present a parallel algorithm for the efficient clustering of objects into groups based on their similarity to each other. The input consists of an n by n distance matrix. This matrix would have a distance ranking for each pair of objects. The smaller the number, the more similar the two objects are to each other. We utilize parallel processors to calculate a hierarchal cluster of these n items based on this matrix. Another advantage of our method is distribution of the large n by n matrix. We have implemented our algorithm and have found it to be scalable both in terms of processing speed and storage.

## 1 Introduction

Dividing similar objects into a smaller number of clusters is of importance in many applications. These include search engines, monitoring of academic performance, biology and wireless networks. We first discuss a number of clustering methods. These include the K-means clustering algorithm and a number of hierarchal clustering methods. Hierarchal clustering methods have advantages over K-means. Unfortunately, Hierarchal Clustering can be expensive. It is for this reason that hierarchal clustering has been avoided in some applications. One example of such an application is the clustering of candidate protein structures into a limited number of groups (Zheng, Gallicchio, Deng, Andrec, & Levy, 2011). If and when clustering is used it is generally K-means clustering. Hierarchal Clustering, itself, is also divided into sub-methods such as simple hierarchal clustering (SHC) and complete hierarchal clustering (CHC). There are a few other methods of hierarchal clustering. While SHC has faster algorithms, CHC, which is more useful, is $O(n^3)$. This can be prohibitive for large n.

We present a parallel algorithm for the efficient clustering of proteins into groups. The input consists of an n by n distance matrix.

## 2 Choice of clustering method

### 2.1 Clustering Methods

Clustering is the generic term for methods of categorizing objects into groups or clusters. There are hierarchal and non-hierarchal methods. There are methods that build clusters top-

down – that is they begin with all item in one large cluster and break down that cluster over a number of iterations. There are also bottom-up or agglomerative algorithms.

Non-hierarchal or partition methods include K-means, Graph Theoretic and other methods.

The non-hierarchal clustering methods in general pre-sets the number of clusters and then fit the n items into those clusters. The name K-means indicates that there is a pre-set number of K clusters. It is an efficient algorithm but does not have the advantages of hierarchal clustering.

Hierarchal clustering methods, on the other hand, first assign every item to its own cluster. For n items, we start, therefore, with n clusters. The algorithm then iterates and combines two of the clusters into one larger cluster. It iterates n times until there is one large cluster. A snapshot is taken after every iteration of what clusters there were at the end of each iteration. This is preserved in an upside down tree (Dendrogram). If our goal, for example, was to divide the data into 10 clusters, we simply look 10 levels down the tree – on that level will be 10 clusters.

Thus, there are numerous advantages of hierarchal clustering. One advantage is that there is no pre-set number of clusters. Another advantage of hierarchal clustering is that it's output is a full tree (also known as a Dendrogram) of clusters. The bottom 'row' of the tree is each of the n items in its own cluster. The second to the bottom row shows n-1 clusters and the top of the tree shows one large cluster. Once finished we can choose any of the output levels of the tree depending on how finely clustered we want it.

Hierarchal Agglomerative Clustering methods include:
1. single-linkage
2. complete linkage
3. Average link
4. Centroids
5. Ward's method

We found that Hierarchal bottom-up agglomerative clustering to be useful in many situations but having a cost of $O(n^3)$. This makes it too expensive to use in many real world applications.

One exception is Single-Linkage hierarchal clustering which can be solved by an algorithm that mimics the Prim's Minimum Spanning Tree Algorithm.

To get a feel for the various hierarchal clustering methods, let's illustrate single-linkage and complete linkage. Complete-linkage clustering tends to produce 'round' clusters whereas single-linkage clustering tends to produce 'long' clusters. The shapes of the other clustering methods are in between.

As an example let's look at the three clusters depicted in figure 1 Lavrenko (2014) Hierarchal Clustering algorithms, in the next iteration, would combine two of these clusters and arrive at a total of two resulting clusters. The difference between the specific hierarchal clustering algorithms is the rule used to decide which of the two clusters to combine.

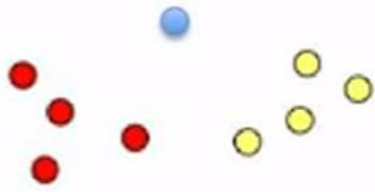

*Figure 1: Example of 3 clusters*

In single linkage we measure the distance between clusters as the distance between their closest members. Single-linkage would combine the red and yellow clusters because the right-most red and left-most yellow are closer than the blue item is to any red or yellow. As you can see we will have an elongated red/yellow cluster where the far yellow and reds may be very different from each other.

In complete linkage we measure the distance between clusters as the distance between their furthest members. Complete linkage would not combine the red and yellow because we will use the distance of the furthest red and furthest yellow which is a larger distance than blue to either yellow or red. It looks like the blue is closer to the furthest yellow (on the right) then to the furthest red (on the bottom-left). Therefore blue would be combines with yellow. In complete linkage, the clusters tend to be circular and closer to each other.

The other schemes produce clusters that are in between; not as elongated as produced by simple-linkage and not as circular as produced by complete-linkage.

Although there exist specialized algorithms for single-linkage (Hendrix et al., 2013), complete linkage tends to give better clusters (not elongated) and is more useful.

For these reasons we employ agglomerative hierarchal complete linkage clustering. This method gives more useful clusters than single-linkage and does not pre-designate a specific number of clusters as in non-hierarchal clustering methods. We use the Lance-Williams update algorithm (Lance & Williams, 1967) and We note that the Lance–Williams algorithm is general and can be used with any of the hierarchal methods listed above; for single-linkage we would suggest more efficient algorithms such as Hendrix (2013).

In the next section we give a brief review of the literature with resources for more study.

## 3   Review of the Literature: Clustering Algorithms

## 3.1   Non-hierarchal methods

There are numerous non-hierarchal clustering methods that begin with a set number of categories and then try to best place items into those categories. K-means, mentioned earlier, is one of the earliest clustering algorithms (Steinhaus, 1956). In addition regression has been used (Gawrysiak, Okoniewski, & Rybinski, 2001). More recently, decision trees methods, which can be useful for categorization into given categories via numerous decision criteria (Liu, Xia, & Yu, 2005) have been used. Kaewchinporn (2011), combined the decision tree method and K-means by used trees to develop weights to be used in K-means. "Random Forest" classification has been developed to address some shortcomings of decision trees. A Random Forest is a group of decision trees where each tree is built via different test data. When it is time to categorize all the trees are used and an average taken. For more information see (Prinzie & Van den Poel, 2008).

## 3.2   Parallel algorithms for Non-hierarchal methods

There has been work on parallel algorithms for some of these methods. Ranka and Sahni (1991) have described a parallel algorithm for K-means on a hypercube. Other parallel algorithms for K-means can be found in (Ahmed, 2014; Dhillon & Modha, 2002; Forman & Zhang, 2000; Judd, McKinley, & Jain, 1998; Kantabutra & Couch, 2000; Li & Fang, 1989; Mathew & Vijayakumar, 2014; Pakhira, 2009; Tsai & Horng, 1999; Y. Zhang, Xiong, Mao, & Ou, 2006; Y.-P. Zhang, Sun, Zhang, & Zhang, 2004; Zhao, Ma, & He, 2009).

The algorithms for these methods are intrinsically more efficient than the algorithms for hierarchal clustering. That is because these methods do not build a hierarchal tree of categories. Instead they pre-suppose a set of categories and then try to fit the input items into those categories.

Some of these methods assume categorical data such as gender and age and some may use numerical data but they all assume a set number of categories.

On the other hand, there are many applications where the clusters differ from each other just as a matter of degree. Take protein structure categorization as an example. We are given n protein chains of exactly the same sequence but they each are folded (their structure or conformation) in a slightly different shape. The chemist or biologist wants to have these proteins structures categorized such that those of a similar structure will be in the same category. It is apparent that a hierarchy of clusters would be useful since there is no way to decide on a pre-set number of clusters. When clustering is used it is typically with K-means due to the cost of hierarchal clustering but it, unfortunately, does not give the best solutions.

## 3.3   Parallel Algorithms for Hierarchal Methods

There has been research for parallel hierarchal clustering. Single-Linkage parallelization was fitted to a hypercube (PRAM) (Olson, 1995). PBIRCH was developed for the BIRCH algorithm (Garg, Mangla, Gupta, & Bhatnagar, 2006; T. Zhang, Ramakrishnan, & Livny, 1997).

Naïve hierarchal clustering, such as the algorithm parallelized in this paper, was presented for an SIMD machine by Ranka and Sahni (Ranka & Sahni, 1991); they focused on a shuffle-exchange network. Li (1990) also describes one for SIMD machines. Parallel algorithms for naïve hierarchal clustering either were for specific architectures or for single-link clustering.

There are a number of thorough surveys of clustering algorithms both serial (Baser & Saini, 2013; Berkhin & others, 2006; Murtagh, 1983; Xu & Wunsch, 2005) and parallel (Kim, 2009; Olson, 1995).

As mentioned in the last section, in this paper we focus specifically on parallelizing the generic Lance-Williams clustering algorithm.

# 4 Lance–William Algorithm

We parallelize the generic Lance William algorithm. This algorithm is able to computer a hierarchal closing dendrogram for many hierarchal clustering schemes.

The Lance-Williams algorithm is iterative. It starts with n clusters and in every iteration combines two clusters together so that after the iteration there are n-1 clusters. The basic algorithm is:

**Lance –William Algorithm**

For k= 1 to n

1. From the distance matrix obtain the minimum distance and set aside the (i,j): $O(n^2)$ time

2. Combine item i and item j into one item thus reducing the number of clusters by 1.

3. Re-calculate new distances between pre-existing clusters and the new combined cluster: O(n) time.

    Use the correct update formula depending upon which method is used; $\boldsymbol{D_{k,i+j} = \alpha_i D_{k,i} + \alpha_j D_{k,j} + \beta D_{i,j} + \gamma |D_{k,i} - D_{k,j}|}$ where α, β and γ depend on the specific method. see Table 1, Lavrenko (2014).

4. Output the current tree level showing the (n-k) clusters.

| Formula used in step 3: $D_{k,i+j} = \alpha_i D_{k,i} + \alpha_j D_{k,j} + \beta D_{i,j} + \gamma |D_{k,i} - D_{k,j}|$ | | | | |
|---|---|---|---|---|
| **Method** | $\alpha_i$ | $\alpha_j$ | $\beta$ | $\gamma$ |
| **Single Linkage** | 0.5 | 0.5 | 0 | -0.5 |
| **Complete Linkage** | 0.5 | 0.5 | 0 | 0.5 |
| **Group Average** | $\dfrac{n_i}{n_i + n_j}$ | $\dfrac{n_j}{n_i + n_j}$ | 0 | 0 |
| **Weighted Average** | 0.5 | 0.5 | 0 | 0 |
| **Centroid** | $\dfrac{n_i}{n_i + n_j}$ | $\dfrac{n_j}{n_i + n_j}$ | $\dfrac{-n_i * n_j}{(n_i + n_j)^2}$ | 0 |
| **Ward** | $\dfrac{n_i+n_k}{(n_i + n_j + n_k)}$ | $\dfrac{n_j+n_k}{(n_i + n_j + n_k)}$ | $\dfrac{-n_k}{(n_i + n_j + n_k)}$ | 0 |

*Table 1: Formulas used in the Lance Williams algorithm for various update methods*

# 5 Parallel Lance Williams

## 5.1 Network

Implementation of the algorithm distributed hierarchical clustering – was compiled and run on Andy (College of Staten Island's High Performance Computing cluster). Andy has 744 cores, Nehalem 2.93 GHz chip, 3 Gb per core ("Home | CUNY High Performance Computing Center," n.d.)

Parallelized RMSD and distributed hierarchical clustering algorithms were implemented using C and MPI (Pacheco, 1997).

RAM is also distributed which makes $\dfrac{n^2-n}{2}$ (upper triangle of n by n matrix) storage feasible since the table is divided up amongst the workstations.

## 5.2 Methodology

We described the general Lance William algorithm above.

To parallelize, we divided up the matrix amongst p processors. Figure 2 shows how this is done. Note that a distance matrix only needs the upper (or lower) diagonal part filled (index i,j will have the same value as index j,i). The number of non-blank cells is divided by p; in this case 28/7 = 4 items per processor. Matrix items are assigned to processors on a row by row basis as in the figure.

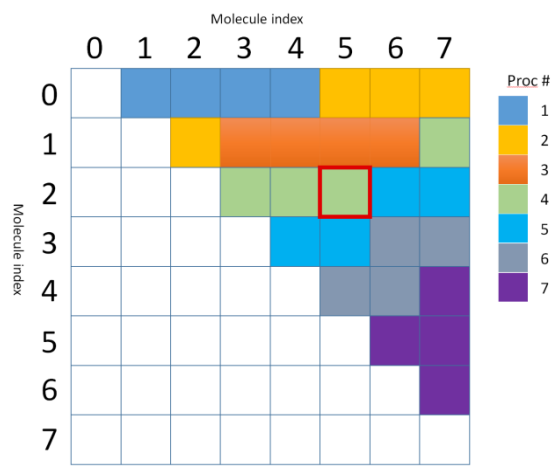

*Figure 2.     Method of dividing the n by n matrix amongst p processors.
In this example n=8 and p=7 and the items are molecules.*

## 5.3     Parallel Lance Williams Algorithm

Each processor $p_m$, $where\ 1 \leq m \leq p$, is sent their portion of the matrix: $\frac{\left(\frac{n^2-n}{2}\right)}{p}$ items.  As the data files were read in from disk they were sent to the processors. The processors are also sent the (i,j) global matrix indices for their data portion. This is especially important for step 6a below.

Divide up matrix amongst the processors.

For k= 1 to n

1. All processors $p_m$, $where\ 1 \leq m \leq p$, simultaneously calculate their local minimum distance from the cells of the distance matrix assigned to them
2. Each $p_m$ broadcasts their local minimum
3. Each $p_m$ now has all p sets (lmin$_m$, $p_m$) where lmin$_m$ is a local minimum and p$_m$ is the corresponding processor.
4. Each processor simultaneously calculates the global minimum (gmin) and the "winning" processor – the process that has the global minimum (i,j) in its part of the matrix. Note that communication is unnecessary at this step since very processor has enough information to calculate the global minimum.
5. The processor that was assigned cell (i,j), which in the example of figure 2 is (2,5), broadcasts that items i and j should be combined into one cluster.
6. Update step: we no longer need columns i and j and rows i and j. The ith column and row will be used for the new combined I,j cluster. The jth column and row will be marked not to be used again.

    Processors that have matrix elements in the ith or jth columns and rows apply the formula:

$$D_{k,i+j} = \alpha_i D_{k,i} + \alpha_j D_{k,j} + \beta D_{i,j} + \gamma |D_{k,i} - D_{k,j}|$$ where α, β and γ depend on the specific method. We used
$$D_{k,i+j} = .5 D_{k,i} + .5 D_{k,j} + .5 |D_{k,i} - D_{k,j}|$$ for complete linkage clustering.

  a. Processors must communicate with each other to obtain any necessary Dk,i from another processor that it needs. Note that the only processors that need to participate in this communication are those that have data items corresponding to rows i or j of the global matrix.
  Processors that have data items from the jth row or jth column send a list of triples (i,j,distance) to all processors that have data items in the ith row or ith column.
  b. The sending processors mark the sent matrix elements as erased not to be used again. The processors that receive the communication apply the formula and fill their previously used ith row and column with the updated combined cluster information. Thus the number of clusters has been reduced by 1.

## 5.4 Complexity Analysis

The naïve Lance Williams algorithm is $O(n^3)$ There is an outer loop that iterated n time where n is the number of items. For each of those the upper diagonal matrix is search a maximum of $\left(\frac{n^2-n}{2}\right)$ times. It is a maximum because after each iteration the size of the matrix is reduced by one row and one column as clusters are combined.

In our parallel version all work is divided evenly amongst the processors. In addition, the matrix elements are evenly divided and stored.

Work per iteration: $\frac{\left(\frac{m^2-m}{2}\right)}{p}$, $where\ 1 \leq m \leq n$; m begins at n and is decremented every iteration

Storage requirement: $\frac{\left(\frac{n^2-n}{2}\right)}{p}$ per distributed unit.

Communication:

- The initial matrix element is sent p separate sends and receives
- At most p broadcasts per iteration (step 2) and
- At most p sends and receives per iteration (step 6a) Note that a subset of processors are involved in this communication.

Thus we have:

Computation: $\left(\frac{\left(\frac{m^2-m}{2}\right)}{p}\right)n\ \ which\ is\ \ O\left(n^3/p\right)$

Communication: $p\ initially, \max of\ 2p*n\ during\ the\ iterations\ which\ is$
$O(p)\ communications, where\ 1\ communication\ is\ a\ send, receive\ pair$

Storage: $\frac{\left(\frac{n^2-n}{2}\right)}{p}$ matrix elements which is $\boldsymbol{O\left(n^2/p\right)}$

## 6  Experimental Results

Figure 2 shows the results of the algorithm as applied to many items. The algorithm was tested many times with varying numbers of items, n, to be clustered. The average of n was 1968.

For each n we ran the program with varying numbers of processors. As we can see from figure 2, speedup is initially linear until about 5 processors. From that point until p=15 the average running time decreased. From  $p$=15 and on the figure show the speed converging. We see that in this case that the optimal number of processors is about p=15. After that, the communication costs amongst processors seem to outweigh any computational gains.

The specific optimum number of processors will grow as the number of items to be clustered grows. The algorithm is scaleable as the problem grows.

Especially noteworthy is that our algorithm cuts down on storage requirements which is a bottleneck for large clustering problems. Use of our algorithm on distributed networks with separate memory spaces seems promising and should make use of hierarchal clustering possible on any distributed network of workstations.

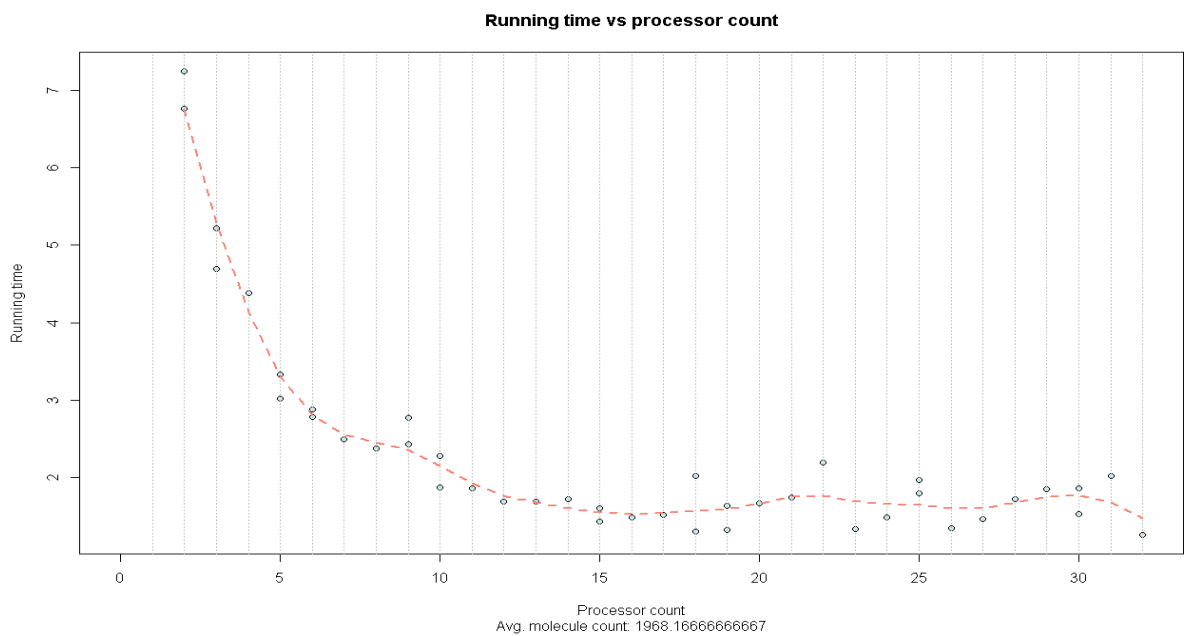

*Figure 2.*     *Running time is shown as function of processor count. The algorithm was run many times and the average number of items is approximately 1968*

# References

Ahmed, M. F. (2014). Parallel Implementation of K-Means on Multi-Core Processors. *Computer Science & Telecommunications*, *41*(1), .

Baser, P., & Saini, J. R. (2013). A Comparative Analysis of Various Clustering Techniques used for Very Large Datasets. *International Journal of Computer Science & Communication Networks*, *3*(5), 271.

Berkhin, P., & others. (2006). A survey of clustering data mining techniques. *Grouping Multidimensional Data*, *25*, 71.

Dhillon, I. S., & Modha, D. S. (2002). A data-clustering algorithm on distributed memory multiprocessors. In *Large-scale parallel data mining* (pp. 245–260). Springer. Retrieved from https://link.springer.com/chapter/10.1007/3-540-46502-2_13

Forman, G., & Zhang, B. (2000). Distributed data clustering can be efficient and exact. *ACM SIGKDD Explorations Newsletter*, *2*(2), 34–38.

Garg, A., Mangla, A., Gupta, N., & Bhatnagar, V. (2006). PBIRCH: A scalable parallel clustering algorithm for incremental data. In *Database Engineering and Applications Symposium, 2006. IDEAS'06. 10th International* (pp. 315–316). IEEE. Retrieved from http://ieeexplore.ieee.org/abstract/document/4041640/

Gawrysiak, P., Okoniewski, M., & Rybinski, H. (2001). Regression—Yet Another Clustering Method. In *Intelligent Information Systems 2001* (pp. 87–95). Springer. Retrieved from http://link.springer.com/chapter/10.1007/978-3-7908-1813-0_8

Hendrix, W., Palsetia, D., Patwary, M. M. A., Agrawal, A., Liao, W., & Choudhary, A. (2013). A scalable algorithm for single-linkage hierarchical clustering on distributed-memory architectures. In *Large-Scale Data Analysis and Visualization (LDAV), 2013 IEEE Symposium on* (pp. 7–13). IEEE. Retrieved from http://ieeexplore.ieee.org/abstract/document/6675153/

Home | CUNY High Performance Computing Center. (n.d.). Retrieved June 7, 2016, from http://www.csi.cuny.edu/cunyhpc/pdf/hpccbrochure.pdf

Judd, D., McKinley, P. K., & Jain, A. K. (1998). Large-scale parallel data clustering. *IEEE Transactions on Pattern Analysis and Machine Intelligence*, *20*(8), 871–876.

Kaewchinporn, C., Vongsuchoto, N., & Srisawat, A. (2011). A combination of decision tree learning and clustering for data classification. In *Computer Science and Software Engineering (JCSSE), 2011 Eighth International Joint Conference on* (pp. 363–367). IEEE. Retrieved from http://ieeexplore.ieee.org/abstract/document/5930148/

Kantabutra, S., & Couch, A. L. (2000). Parallel K-means clustering algorithm on NOWs. *NECTEC Technical Journal*, *1*(6), 243–247.

Kim, W. (2009). Parallel clustering algorithms: survey. *Parallel Algorithms, Spring*. Retrieved from https://pdfs.semanticscholar.org/9b6c/a50781bcbcc5613f7b120b8295cd3f6ed933.pdf

Lance, G. N., & Williams, W. T. (1967). A general theory of classificatory sorting strategies: II. Clustering systems. *The Computer Journal*, *10*(3), 271–277.

Li, X., & Fang, Z. (1989). Parallel clustering algorithms. *Parallel Computing*, *11*(3), 275–290.

Liu, B., Xia, Y., & Yu, P. S. (2005). Clustering Via Decision Tree Construction. In *Foundations and advances in data mining* (pp. 97–124). Springer. Retrieved from http://link.springer.com/chapter/10.1007/11362197_5

Mathew, J., & Vijayakumar, R. (2014). Scalable parallel clustering approach for large data using parallel K means and firefly algorithms. In *High Performance Computing and Applications (ICHPCA), 2014 International Conference on* (pp. 1–8). IEEE. Retrieved from http://ieeexplore.ieee.org/abstract/document/7045322/

Murtagh, F. (1983). A survey of recent advances in hierarchical clustering algorithms. *The Computer Journal*, *26*(4), 354–359.

Olson, C. F. (1995). Parallel algorithms for hierarchical clustering. *Parallel Computing*, *21*(8), 1313–1325.

Pacheco, P. S. (1997). *Parallel programming with MPI*. Morgan Kaufmann. Retrieved from https://books.google.com/books?hl=en&lr=&id=GufgfWSHt28C&oi=fnd&pg=PR7&dq=Parallel+Programming+with+MPI&ots=6mPq23mPpc&sig=F2houNMM9_xRcnWdTFhWuBTShHQ

Pakhira, M. K. (2009). Clustering large databases in distributed environment. In *Advance Computing Conference, 2009. IACC 2009. IEEE International* (pp. 351–358). IEEE. Retrieved from http://ieeexplore.ieee.org/abstract/document/4809035/

Prinzie, A., & Van den Poel, D. (2008). Random forests for multiclass classification: Random multinomial logit. *Expert Systems with Applications*, *34*(3), 1721–1732.

Ranka, S., & Sahni, S. (1991). Clustering on a hypercube multicomputer. *IEEE Transactions on Parallel and Distributed Systems*, *2*(2), 129–137.

Steinhaus, H. (1956). Sur la division des corp materiels en parties. *Bull. Acad. Polon. Sci*, *1*(804), 801.

Tsai, H.-R., & Horng, S.-J. (1999). Optimal parallel clustering algorithms on a reconfigurable array of processors with wider bus networks. *Image and Vision Computing*, *17*(13), 925–936.

Victor Lavrenko. (2014). *Hierarchical Clustering 4: the Lance-Williams algorithm*. Retrieved from https://www.youtube.com/watch?v=aXsaFNVzzfI

Xu, R., & Wunsch, D. (2005). Survey of clustering algorithms. *IEEE Transactions on Neural Networks*, *16*(3), 645–678.

Zhang, T., Ramakrishnan, R., & Livny, M. (1997). BIRCH: A new data clustering algorithm and its applications. *Data Mining and Knowledge Discovery*, *1*(2), 141–182.

Zhang, Y., Xiong, Z., Mao, J., & Ou, L. (2006). The study of parallel k-means algorithm. In *Intelligent Control and Automation, 2006. WCICA 2006. The Sixth World Congress on* (Vol. 2, pp. 5868–5871). IEEE. Retrieved from http://ieeexplore.ieee.org/abstract/document/1714203/

Zhang, Y.-P., Sun, J.-Z., Zhang, Y., & Zhang, X. (2004). Parallel implementation of CLARANS using PVM. In *Machine Learning and Cybernetics, 2004. Proceedings of 2004 International Conference on* (Vol. 3, pp. 1646–1649). IEEE. Retrieved from http://ieeexplore.ieee.org/abstract/document/1382039/

Zhao, W., Ma, H., & He, Q. (2009). Parallel k-means clustering based on mapreduce. In *IEEE International Conference on Cloud Computing* (pp. 674–679). Springer. Retrieved from http://link.springer.com/10.1007/978-3-642-10665-1_71

Zheng, W., Gallicchio, E., Deng, N., Andrec, M., & Levy, R. M. (2011). Kinetic network study of the diversity and temperature dependence of trp-cage folding pathways: Combining transition path theory with stochastic simulations. *The Journal of Physical Chemistry B*, *115*(6), 1512–1523.